\documentstyle[aps,epsf,preprint]{revtex}
\newcommand{\bb}{\begin{equation}}
\newcommand{\en}{\end{equation}}
\draft
\begin{document}
\date{\today}
%
%
%
%

\title{Charge Fluctuations and Counterion Condensation}
\author{A.W.C. Lau$^{1}$, D.B. Lukatsky$^{3}$, P. Pincus$^{2}$, and
S.A. Safran$^{3}$}
\address{$^{1}$ Department of Physics and Astronomy, University of Pennsylvania, Philadelphia, PA 19104\\
$^{2}$ Materials Research Laboratory, University of California, Santa Barbara, CA 93106--9530 \\
$^{3}$ Department of Materials and Interfaces, Weizmann Institute, Rehovot, 76100 Israel}
\maketitle

\begin{abstract}
We predict a condensation phenomenon in an overall neutral system,
consisting of a single charged plate and its oppositely charged counterions.
Based on the ``two-fluid'' model, in which the counterions are divided
into a ``free'' and a ``condensed'' fraction, we argue that
for high surface charge, fluctuations can
lead to a phase transition in which a large fraction of
counterions is condensed.  Furthermore, we show that depending on
the valence, the condensation is either a first-order or a smooth
transition.
 \end{abstract}

\narrowtext

\section{Introduction}
\label{sec:introduction}

Electrostatic interactions control the structure, phase behavior,
and function of macroions in aqueous solutions\cite{nature}.  The
macroions may be charged membranes, stiff polyelectrolytes such as
DNA, or charged colloidal particles. The fundamental description
of these charged systems has been the Poisson-Boltzmann (PB)
theory.  However, it ignores fluctuations and correlations, which
are important for the cases of low temperatures, highly charged
surfaces, or multivalent counterions. These fluctuation and
correlation effects, which have been the focus of recent
theoretical efforts, may drastically alter the mean-field picture
of PB theory \cite{fluct,netz,shklovskii,attractionT}. For
example, one surprising effect\cite{attractionT} is the {\em
attraction} between two highly charged macroions, as observed in
experiments\cite{attractionE} and in simulations\cite{attractionS}.
In this paper, we argue that
correlation effects may lead to condensation of counterions
onto an oppositely charged plate, whose surface charge becomes
effectively renormalized.  In particular, the counterion valence plays an
interesting role: for $Z > Z_c \sim 1.62$ for typical system parameters (see below),
we find a first-order phase transition in which a large fraction
of the counterions is {\em condensed}, while for $Z < Z_c$ the condensation
proceeds smoothly, implying that monovalent and divalent counterions exhibit
qualitatively distinct behavior.  This is in contrast with more
familiar theories of counterion condensation\cite{manning},
{\em e.g.} Manning condensation for charged rods, where the effective
charge is continuously modified by the valence.

Recall that for a single plate of charge density
$\sigma({\bf x}) = \sigma_0 \delta(z)$ immersed in an aqueous solution
of dielectric constant $\epsilon$,
containing point-like counterions of charge $- Z e$ on both sides of the
plate, PB theory predicts that the counterion density\cite{nature}\bb
c(z) =  {1   \over 2 \pi Z^2 l_B\left ( |z| + \lambda \right )^2},
\label{scd}
\en
decays to zero algebraically with a characteristic length
$\lambda \equiv  { e /(\pi l_B Z \sigma_0)}$, where
$l_B \equiv {e^{2} \over \epsilon k_{B}T} \approx 7\,$\AA$\,\,$
is the Bjerrum length in water at room temperature,
$k_B$ is the Boltzmann constant, and $T$ is the temperature.
This Gouy-Chapman (GC) length $\lambda$ defines a sheath near the
charged surface within which most of the counterions are confined.
Typically, it is on the order of few Angstroms
for $\sigma_0 \sim e /100\,\,$\AA$^{-2}$.  Note that Eq. (\ref{scd})
implies that at {\em zero} temperature all of the counterions would
collapse onto the charged plane.
However, for high surface charge (or low temperature) $Z^2 {l}_B \gg \lambda$,
fluctuation and correlation corrections can become so large that
the solution Eq. (\ref{scd}) to the PB equation is no longer
valid\cite{netz}. Therefore, we might expect a quantitative
deviation from the conclusion above.  Indeed, as pointed out by Netz {\em et
al.}\cite{netz}, a perturbative expansion about the
PB solution breaks down in this regime, as indicated by an unphysical
(negative) counterion density in the one-loop approximation.
Motivated by these observations, we propose a {\em two-fluid} model in
which the counterions are divided into a {\em free} and a
condensate fraction. The {\em free} counterions have the usual 3D
spatial distribution, while the {\em condensed} counterions are
confined to the two-dimensional charged plane, with a mean (2D) density $n_c$.
We treat the fraction of 2D condensed counterions $\tau \equiv Ze n_c /\sigma_0$
as a variational parameter, which is determined self-consistently
by minimizing the total free energy of the system.

It may be useful to illustrate the essential physics first by a simple picture.
In the spirit of the two-fluid model, the 2D condensed counterions
partially neutralize the charged plate, effectively reducing the surface
charge density from $\sigma_0$ to $e n_R = \sigma_0 - Ze n_c$, where $n_c$
is their surface (2D) density.  The free counterions can be modeled
as a 3D ideal gas confined to a slab of thickness
$\lambda_R \equiv  1 /(\pi l_B Z n_R)$.  At the Debye-H\"{u}ckel level,
the free energy per unit area for the condensed
counterions $f_{2d}(n_c)$ can be written as\cite{2Dfree}\bb
\beta f_{2d}(n_c) = n_c\left \{ \ln[ n_c\,a^2] - 1 \right \} + { 1 \over
2}\int {d^{2} {\bf q} \over (2 \pi)^{2}} \left \{ \ln \left [ 1 +
\frac{1}{q\lambda_{D}} \right ] - \frac{1}{q\lambda_{D}} \right
\}, \label{2D}
\en
where $\beta^{-1} = k_B T$ and $\lambda_{D} =
1/(2 \pi l_B Z^2 n_{c})$ is the 2D screening length.  The first
term in Eq. (\ref{2D}) is the entropy and the second term arises
from the 2D fluctuations.  Note that the latter term is
logarithmically divergent, which may be regularized by a
microscopic cut-off $q_c \sim 2 \pi /a$, yielding $\beta \Delta f_{2d}(n_c)
\simeq  -{1 \over 8 \pi \lambda_{D}^2}\ln(2\pi \lambda_{D}/a)$.
The free energy of the free counterions $f_{3d}(n_c)$ consists of
the entropy of a confined 3D ideal gas and the fluctuation free
energy. The latter term may be estimated by using the fluctuation
contribution to the free energy density from the 3D
Debye-H\"{u}ckel theory\cite{landau} and multiplying it by the
thickness of the slab $\lambda_R$:\bb
\beta f_{3d}(n_c) \approx c\, \lambda_R  \left \{ \ln[ c\,a^3] - 1
\right \}  - {\kappa_s^3 \over 12 \pi}\,\lambda_R,
\label{app3dfree}
\en where $c = n_R / (Z\lambda_R)$ average (3D)
concentration of the free counterions and $ \kappa_s^2 \equiv 4
\pi Z^2 l_B c$ is the $3D$ screening length.  Note that the second term scales as
$\sim - \,\lambda_R^{-2}$.  This simple picture to estimate $f_{3d}(n_c)$ contains all the
qualitative physics\cite{note3}, which follow from the more precise analysis presented below.
The total free energy in the two-fluid model is $f(\tau) = f_{2d}(\tau) + f_{3d}(\tau)$.
Minimizing $f(n_c)$ to find $n_c$, we obtain \bb
1 + \tau g \ln \left ( {\pi \over \tau \theta g} \right ) - \ln \left [ \tau \over ( 1 - \tau )^2 \theta g \right ]
- {4 \over 3}\,g (1 - \tau) = 0,
\label{eqnofstate}
\en
where the three dimensionless parameters:
the order parameter $\tau \equiv Ze n_c /\sigma_0$,
the coupling constant $g \equiv Z^2 {l}_B / \lambda$, (where
$\lambda$ is the {\em bare} GC length), and the reduced
temperature $\theta \equiv {a /( Z^2 l_B)}$, completely determine
the equilibrium state of the system.  It is straightforward to obtain
the asymptotic solutions of Eq. (\ref{eqnofstate}) corresponding
to the uncondensed, $\tau_1 \ll 1$,
and condensed, $\tau_2 \approx 1$, state of
the counterions:  $\tau_1  \simeq g\,\theta \exp \left[ 1-\frac{4}{3}\,g\right]$
and $\tau_2 \simeq 1 - \left[ \pi \exp (1)\right] ^{-1/2}\left( \frac{g \theta }{\pi }
\right) ^{g - 1 \over 2}$.  For weak couplings, $g \ll 1$, $\tau_1$
is the only consistent solution.
On the other hand, for large coupling $g \gg 1$, where fluctuation
free energies dominate the system, $\tau_1$ and $\tau_2$ are both consistent
solutions for small $\theta$, and a first-order transition takes place
when $f(\tau_1) = f(\tau_2)$.  Thus, a large fraction of counterions is
condensed if $g$ exceeds some threshold value $g > g_0$. For an estimate, taking
$\theta = 0.02$ (divalent counterions at room temperature) we find
$g_0 \sim 1.757$, corresponding to a surface charge of
$\sigma_0 \sim e/10$ nm$^{-2}$.

We emphasize that although there is a close analogy between our approach and
the more familiar theory of counterion condensation, {\em e.g.}
Manning condensation\cite{manning}, the counterion condensation in
our model has a different physical origin arising from charge fluctuations.
In Manning condensation, the competition between entropy and electrostatics
leads to an electrostatic potential at large distances that is independent
of the charge density of the rod above the Manning threshold\cite{manning}.
In this sense, for the geometry of a charge plate, counterions are
always "Manning condensed" at the PB level\cite{Zimm}.  On the other hand,
in our model, we take one step further by showing that when correlation effects
are taken into account, a finite fraction of the counterions
is condensed to form a 2D Coulomb gas onto the charged plate.
This paper is organized as follows:  In Sec. \ref{sec:twofluid},
we present in detail the two-fluid model and construct the total
free energy of the system.  In Sec. \ref{sec:results}
we present the central results of this paper, followed by
an extensive discussion.

\section{Counterion Free Energy in The ``Two-Fluid'' Model}
\label{sec:twofluid}

To study the condensation more rigorously, we compute total free energy by
mapping the problem into a field theory.  Consider an overall neutral system
consisting of counterions and an oppositely charged surface immersed in an aqueous
solution.  The surface charge density on the plate is $\sigma_0 = e n_0$.
We model the aqueous solution with a uniform dielectric constant
$\epsilon$.  This simplification allows us to study fluctuation and
correlation effects analytically.  In the spirit of the ``two-fluid'' model,
we divide the counterions into a ``condensed'' and a ``free'' fraction.
The condensed counterions are allowed to move only on the charged surface, while
the free counterions distribute in the space on both sides of the plate.
The electrostatic free energy for the whole system may be written as\begin{eqnarray}
\beta F_{el} &=& \int d^2{\bf r}\, n_c({\bf r})
\left \{ \ln \left [ n_c({\bf r}) a^2 \right ] -  1 \right \}
+  \int d^3{\bf x}\, \rho({\bf x})
\left \{ \ln \left [ \rho({\bf x}) a^3 \right ] -  1 \right \} \nonumber \\
&+&  {Z^2l_B\over 2}\, \int d^3{\bf x} \int d^3{\bf x}'\,
{ n_c({\bf r})\delta(z)\,n_c({\bf r}')\delta(z') \over | {\bf
x} - {\bf x}' | }  + {Z^2 l_B \over 2}\,
\int d^3{\bf x} \int d^3{\bf x}'\,{ \rho({\bf x}) \rho({\bf x}') \over | {\bf
x} - {\bf x}' | } \nonumber \\
&+& {Z l_B } \int d^3{\bf x} \int d^3{\bf x}'\,
{ n_c({\bf r})\delta(z) [Z \rho({\bf x}') - n_f({\bf x}')] \over | {\bf x} - {\bf x}' | }
-  {Z l_B }\int d^3{\bf x} \int d^3{\bf x}'\,
{  \rho({\bf x})\,n_f({\bf x}') \over | {\bf x} - {\bf x}' |} \nonumber \\
&+& {l_B \over 2}\int d^3{\bf x} \int d^3{\bf x}'\,
{  n_f({\bf x}) n_f({\bf x}') \over | {\bf x} - {\bf x}' |},
\label{2dfree1}
\end{eqnarray}
where $a$ is the molecular size of the counterions,
$l_B = e^2 / (\epsilon k_B T)$ is the Bjerrum length, $Z$
is the valence of the counterions, ${\bf r}$ is the in-plane position vector,
and ${\bf x} = ( {\bf r},z)$.  The first two terms in Eq. (\ref{2dfree1}) are
the two-dimensional entropy for the
condensate and three-dimensional entropy for the ``free'' counterions, respectively, and
the other terms  represent the electrostatic interactions of counterions in the system.
In Eq. (\ref{2dfree1}), the two-dimensional density of the condensed counterions
is denoted by $n_c({\bf r})$, the ``free'' counterions with 3D density by $\rho({\bf x})$, and
the external fixed charges arising from the surface by $n_f({\bf x}) =
n_0 \delta(z)$.  Within the Gaussian fluctuation approximation,
we consider the spatial dependent fluctuations of the 2D density
of condensed counterions about a uniform mean:
$n_c({\bf r}) = n_c + \delta n_c({\bf r})$, and expand
Eq. (\ref{2dfree1}) to second order in $\delta n_c({\bf r})$:
\begin{eqnarray}
\beta F_{el} &=&  n_c \left \{ \ln \left [ n_c a^2
\right ] -  1 \right \} {\cal A}\,+ \int d^3{\bf x}\, \rho({\bf x}) \left \{ \ln \left [ \rho({\bf x}) a^3
\right ] -  1 \right \} \nonumber \\
&+& {Z^2 l_B \over 2}\, \int d^3{\bf x}
\int d^3{\bf x}'\,{ \rho({\bf x}) \rho({\bf x}') \over | {\bf
x} - {\bf x}' | }- {Z l_B } \int d^3{\bf x} \int d^3{\bf x}'\,
{  \rho({\bf x})n_R({\bf x}') \over | {\bf x} - {\bf x}' |}  \nonumber \\
&+& { 1 \over 2}  \int d^2{\bf r} \int d^2{\bf r}'\,
\left [ {Z^2 l_B  \over  | {\bf r} - {\bf r}' |} + {\delta( {\bf r} - {\bf r}') \over n_c}
\right ] \delta n_c({\bf r}) \delta n_c({\bf r}')\nonumber \\
&+& {Z l_B } \int d^3{\bf x} \int d^3{\bf x}'\,
{ \delta n_c({\bf r})\delta(z) J({\bf x}')\over | {\bf x} - {\bf x}' | }
+ {l_B \over 2}\int d^3{\bf x} \int d^3{\bf x}'\,
{  n_R({\bf x}) n_R({\bf x}') \over | {\bf x} - {\bf x}' |} + {\cal O}[\delta n_c({\bf r})]^3,
\nonumber
\end{eqnarray}
where ${\cal A}$ is the area of the plane, $J({\bf x})
\equiv Z \rho({\bf x}') - n_R({\bf x})$, and $n_R({\bf x}) = n_f({\bf x}) -
Z n_c \delta( z )$.  Note that $J({\bf x})$ is linearly coupled to
$\delta n_c({\bf r})$ in the above equation.  Summing over all the 2D fluctuations
of the condensed counterions, {\em i.e.}\[
e^{ - \beta {\cal H}_e}  =
\int {\cal D}\delta n_c({\bf r})  e^{-\beta F_{el}},
\]
we obtain two terms in the effective free energy:  ${\cal H}_e = F_{2d} + {\cal H}_{3d}$.
The first term $F_{2d}$ is the free energy associated with the condensed
counterions which can be written as\bb
\beta F_{2d} =  n_c\left \{ \ln[ n_c\,a^2] - 1 \right \} {\cal A} +
{1 \over 2} \ln \det \hat{{\bf K}}_{2d} - {1 \over 2} \ln \det[ -\nabla_{\bf x}^2],
\label{2DExp}
\en
where $\hat{{\bf K}}_{2d}({\bf x}, {\bf y}) \equiv
\left [ - \nabla_{{\bf x}}^2 + {2 \over
\lambda_D}\, \delta(z) \,\right ]\delta( {\bf x} - {\bf
y})$ is the 2D Debye-H\"{u}ckel operator and $\lambda_{D} =  1/(2\pi Z^2 l_B n_{c})$
is the Debye screening length in 2-D.  The first term in Eq. (\ref{2DExp}) is the entropy
and the second term arises from the 2D charge-fluctuations.  Note that
this fluctuation term can be evaluated analytically\cite{2Dfree},
with the result quoted in Eq. (\ref{2D}):\[
\beta f_{2d}(n_c) = n_c\left \{ \ln[ n_c\,a^2] - 1 \right \} + { 1 \over
2}\int {d^{2} {\bf q} \over (2 \pi)^{2}} \left \{ \ln \left [ 1 +
\frac{1}{q\lambda_{D}} \right ] - \frac{1}{q\lambda_{D}} \right
\}.
\]

The second term ${\cal H}_{3d}$ is the electrostatic free energy
for the ``free'' counterions, taking into account of the presence
of the fluctuating condensate; to within an additive constant, it may be written as
\begin{eqnarray} \beta {\cal H}_{3d} &= &\int d^3{\bf x}\,
\rho({\bf x}) \left \{ \ln \left [ \rho({\bf x}) a^3 \right ] -  1
\right \}  + {1 \over 2}\,\int d^3{\bf x} \int d^3{\bf
x}'\,\rho({\bf x})
G_{2d}({\bf x}, {\bf x}')\rho({\bf x}') \nonumber \\
&-&  \int d^3{\bf x}\,\phi({\bf x})\,\rho({\bf x}),
\label{energy}
\end{eqnarray}
where $\phi({\bf x}) \equiv \int d^3{\bf x}'\,Z^{-1}\, G_{2d}({\bf
x}, {\bf x}')\,n_R({\bf x}')$ is the ``renormalized'' external
field arising from the charged plate.  From Eq. (\ref{energy}), we
can see that the presence of the condensate modifies the
electrostatics of the free counterions in two ways.  First, the
condensate partially neutralizes the charged surface, effectively
reducing the surface charge density from $e n_0$ to $e n_R = e(n_0
- Z n_c)$.  Second, their fluctuations renormalize the
electrostatic interaction of the system; thus, instead of the
usual Coulomb potential, the free counterions and the charged
surfaces interact via the interaction $G_{2d}({\bf x},{\bf x}')$,
which is the inverse (the Green's function) of the 2D
Debye-H\"{u}ckel operator $\hat{{\bf K}}_{2d}$\cite{2Dscreen}:\bb
\left [ - \nabla_{{\bf x}}^2 + {2 \over \lambda_D}\,\delta(z)
\right ] G_{2d}({\bf x}, {\bf x}') =  4 \pi l_B Z^2 \delta({\bf x}
- {\bf x}'), \label{2dDebye} \en where  the second term in the
bracket takes the fluctuating 2D ``condensate'' into account.
Hence, in the limit $n_c \rightarrow 0$ or $\lambda_{D}
\rightarrow \infty$, $G_{2d}({\bf x},{\bf x}')$ reduces to the
usual Coulomb interaction $G_{0}({\bf x},{\bf x}') = 4 \pi l_B Z^2
/ |{\bf x}-{\bf x}'|$.

After a Hubbard-Stratonovich transformation\cite{hubstr},
the grand canonical partition function for the free counterions
can be mapped onto a functional integral representation:
${\cal Z}_{\mu}[\phi] = {\cal N}_0\,\int {\cal D}\psi\,
e^{- {\cal S}[\psi,\phi]}$ with the effective Hamiltonian\cite{samu}\bb
{\cal S}[\psi,\phi] =  {1 \over 4 \pi l_B Z^2 }\,\int d^3{\bf x}\,\left \{\,
{1 \over 2}\,\psi({\bf x}) [\, - \nabla^2 \,] \psi({\bf x}) +
{1 \over \lambda_D}\,\delta(z)\,
[\psi({\bf x})]^2 - \kappa^2\,e^{i \psi({\bf x}) + \phi({\bf x})\,} \right \},
\label{action}
\en
where $\psi({\bf x})$ is the fluctuating field, $\kappa^2 = 4 \pi l_B Z^2\,e^{\mu}/a^3 $,
$\mu$ is the chemical potential, and ${\cal N}_0^{-2} \equiv \det \hat{{\bf K}}_{2d}$
is the normalization factor.  The minimum of the effective Hamiltonian,
given by $\left. { \delta {\cal S} \over
\delta \psi({\bf x}) }\right |_{\psi =\psi_0} = 0$,
defines the saddle-point equation for $\psi_0({\bf x})$, which reads\bb
\nabla^2 \varphi( {\bf x}) + \kappa^2 e^{- \varphi( {\bf x})}
= { 4 \pi l_B Z n_R}\,\delta(z) +
{ 2 \over \lambda_D }\,\delta(z)\,\varphi({\bf x})
\label{saddle}
\en
in terms of the mean-field potential
$\varphi({\bf x}) = - i \psi_{0}({\bf x}) - \phi({\bf x})$.
The solution to Eq. (\ref{saddle}) is $\varphi({\bf x}) = 2 \ln \left
( 1 + { \kappa |z| \over \sqrt{2} } \right )$,
which satisfies the boundary conditions: $i)$ $\varphi(0)=0$ and
$ii)$ $\left. {d\varphi \over dz} \right |_{z=0} = {2  \pi l_B Z n_R }$,
with $\kappa = 2 \pi l_B Z n_R /\sqrt{2}$.
Thus, at the mean-field level, the distribution of the free counterions\[
\rho_0({\bf x}) \equiv  \kappa^2\,e^{-\varphi({\bf x})} / 4 \pi l_B Z^2
= {1  \over \,2 \pi l_B Z^2 ( |z| + \lambda_R)^2\,}
\]
has exactly the same form as the PB
distribution Eq. (\ref{scd}), but with a {\em renormalized} GC length
$\lambda_R \equiv \sqrt{2}/ \kappa = 1 /( \pi l_B Z n_R)$.
To obtain the mean-field free energy of the free counterions $F_0(n_R)$,
we note that it is related
to the Gibbs potential $\Gamma_0[\phi] \equiv  {\cal S}[\psi_0, \phi]$ by
a Legendre transformation: $ F_0(n_R) =
\Gamma_0[\phi] + \mu \int d^3{\bf x}\,
\rho_0({\bf x})$. Solving for the chemical potential $\mu$ from its
definition: $\mu = \ln \left ( {n_R a^3 \over 2 Z \lambda_R} \right )$
and using the mean-field solution $\varphi({\bf x})$, we
find\bb
\beta F_{0}(n_R)/{\cal A} = {n_R \over Z} \ln \left ( {n_R\,a^3 \over 2 Z \lambda_R} \right )
- {n_R \over Z},
\label{meanfree}
\en
where ${\cal A}$ is the area of the charged plane.
Note that $F_{0}(n_R)$ has the form of an ideal gas entropy
of a gas with concentration $n_R /(Z \lambda_R)$
confined to a slab of thickness $\lambda_R$, the renormalized GC length.

Next, to capture correlation effects, we must also include the fluctuations of the
free counterions, thereby treating the ``free'' and ``condensed''
counterions on the same level.  To this end, we expand the action ${\cal S}[\psi,\phi]$
about the saddle-point $\psi_0({\bf x})$ to second order in $\Delta \psi({\bf x}) =
\psi({\bf x}) -\psi_0({\bf x})$\bb
{\cal S}[\phi,\psi] = {\cal S}[\phi,\psi_0]
+ {1 \over 2}\,\int d^3{\bf x}\,\int d^3{\bf y}\,
\Delta\psi({\bf x})\,\hat{{\bf K}}_{3d}({\bf x}, {\bf y})\,\Delta\psi({\bf y})+ \cdots,
\en
where the differential operator\bb
\hat{{\bf K}}_{3d}({\bf x}, {\bf y}) \equiv
\left [ - \nabla_{{\bf x}}^2\,+\,{2 \over \lambda_D}\,\delta(z)\,+ \,
{ 2 \over ( |z| + \lambda_R)^2}\, \right ] \delta( {\bf x} - {\bf y}),
\label{K}
\en
is the second variation of the action ${\cal S}[\psi,\phi]$.  Note
that the linear term in $\Delta \psi({\bf x})$ does not contribute to the expansion
since $\psi_0({\bf x})$ satisfies the saddle-point equation Eq. (\ref{saddle}).
Performing the Gaussian integrals in the functional integral, we obtain an
expression for the change in the free energy due to fluctuations of the free
counterions:\bb
\beta \Delta F_{3d}= {1 \over 2} \ln \det \hat{{\bf K}}_{3d}  -
{1 \over 2} \ln \det \hat{{\bf K}}_{2d},
\label{3D}
\en
where the second term comes from the normalization factor ${\cal N}_0$.
To evaluate $ \beta \Delta {F}_{3d}$ explicitly, we first
differentiate it with respect to ${l}_{B}$ by making use of the identity
$\delta \ln \det \hat{\bf X} = {\mbox{Tr}}\,\hat{\bf X}^{-1}\,
\delta\,\hat{\bf X}$ to obtain\bb
 4 \pi l_B Z^2{\,\partial  \beta \Delta {F}_{3d} \over \partial {l}_B }
= - {\partial \lambda_R \over \partial {l}_B}
\int d^{3}{\bf x}\,{2 G_{3d}({\bf x},{\bf x}) \over ( |z| + \lambda_R)^3}\,
- {\partial \lambda_D \over \partial {l}_B}
\int d^{3}{\bf x}\,\left
[ G_{3d}({\bf x},{\bf x}) - G_{2d}({\bf x},{\bf x}) \right ]\,{\delta(z) \over \lambda_D^2} ,
\label{derivative}
\en
where $G_{2d}({\bf x}, {\bf x})= \int {d^{2} {\bf q} \over
(2 \pi )^2 }\, {2 \pi l_B Z^2 \over  q} \left [ 1  - { e^{-2 q |z|} \over 1 + q
\lambda_D} \right ]$ and $G_{3d}({\bf x},{\bf x}')$
is the Green's function for the 3D free counterions. It satisfies\bb
\left [\, - \nabla_{{\bf x}}^2 + {2 \over \lambda_D}\,\delta(z) +
{ 2 \over ( |z| + \lambda_R)^2}\,\right ]
G_{3d}({\bf x},{\bf x}')= 4 \pi l_B Z^2 \delta( {\bf x} - {\bf x}'),
\en
which can be solved to yield:
\begin{eqnarray}
G_{3d}({\bf x},{\bf x}) = \int {d^{2} {\bf q} \over
(2 \pi )^2 }\, {2 \pi l_B Z^2 \over q} \left \{\,1
\vphantom{{ \gamma\,(q \lambda_R)^3\,
e^{-2q|z|} \left [ 1 + { 1 \over q\,( |z| + \lambda_R ) } \right ]^2
\over [ 1 + q \lambda_R + (q \lambda_R)^2] \,[( 1 + q \lambda_R )( 1 + \gamma)
+ (q \lambda_R)^2] }} \right.
&-& { 1 \over q^2 ( |z| + \lambda_R)^2 }
+ { e^{-2q|z|} \left [ 1 + { 1 \over q\,( |z| + \lambda_R ) } \right ]^2
\over ( 1 + q \lambda_R ) [ 1 + q \lambda_R + (q \lambda_R)^2] } \nonumber \\
&-& \left. { \gamma\,(q \lambda_R)^3\,
e^{-2q|z|} \left [ 1 + { 1 \over q\,( |z| + \lambda_R ) } \right ]^2
\over [ 1 + q \lambda_R + (q \lambda_R)^2] \,[( 1 + q \lambda_R )( 1 + \gamma)
+ (q \lambda_R)^2] }\,
\right \},
\label{3dgreene}
\end{eqnarray}
where $\gamma \equiv {\lambda_R / \lambda_D} = 2\,\tau /( 1 - \tau)$.
Note that the first term in $G_{3d}({\bf x},{\bf x})$ is just the
Coulomb self-energy $G_0({\bf 0}) = \int {d^{2} {\bf q} \over
(2 \pi )^2 }\, {2 \pi l_B Z^2 \over q}$, which must be subtracted.
Inserting $G_{2d}({\bf x},{\bf x})$ and $G_{3d}({\bf x},{\bf x})$
into Eq. (\ref{derivative}), we obtain \bb
{ 1 \over {\cal A}}{\partial \beta \Delta {F}_{3d}
\over \partial {l}_B } = { {\cal I}_1(\gamma) \over 4 \pi \lambda_R^3 }
\,{\partial \lambda_R \over \partial {l}_B} +
{ {\cal I}_2(\gamma)/\gamma \over 4 \pi \lambda_D^3 }
\,{\partial \lambda_D \over \partial {l}_B} + (\mbox{self-energy}),
\en
where the functions ${\cal I}_{1,2}(\gamma)$
are given by\begin{eqnarray}
{\cal I}_1(\gamma) &=& { 1\over 2} \ln ( 1 + \gamma) + 3 \left |\,
\sqrt{ { 1 + \gamma \over  3 - \gamma} }\,
\tan^{-1} \sqrt{ {  3 - \gamma \over 1 + \gamma } }\, \right |,
\nonumber \\
{\cal I}_2(\gamma) &=& {\gamma \over 2} \ln { \gamma^2 \over 1 + \gamma}
+ ( 2- \gamma) \left |\,
\sqrt{ { 1 + \gamma \over  3 - \gamma} }\,
\tan^{-1} \sqrt{ {  3 - \gamma \over 1 + \gamma } } \,\right |.
\nonumber
\end{eqnarray}
Because ${\cal I}_{1,2}(\gamma)$ are independent of $l_B$, we can
integrate Eq. (\ref{derivative}) back to obtain $ \beta \Delta {F}_{3d}$;
thus, the total free energy per unit area for the free
counterions is\bb
\beta f_{3d}(\tau) =
{n_R \over Z} \ln \left ( {n_R\,a^3 \over 2 Z \lambda_R} \right )
- {n_R \over Z} - { {\cal I}_1(\gamma) \over 8 \pi \lambda_R^2 }
- { {\cal I}_2(\gamma) \over 8 \pi \lambda_D \lambda_R }.
\label{3dfree}
\en

Incidentally, in the limit of vanishing density of the condensed counterions,
$n_c \rightarrow 0$ (or $\lambda_R
\rightarrow \lambda$), ${\cal I}_1(0) = {\pi \over \sqrt{3}}$, and
we obtain the fluctuation correction to the mean-field PB free energy:
$\Delta f_{pb} = -\,k_B T /( 8 \sqrt{3}\,\lambda^2)$.  This result may be understood
physically as follows.  According to PB theory,
the counterions are confined to a slab of thickness $\lambda$,
and thus may be considered as an ideal gas with a 3D concentration of
$c \sim n_0/ \lambda$.  This implies that the inverse of the
3D ``screening'' length is $ \kappa_s \sim \sqrt{c\,l_B} \sim 1/\lambda$.
Using the 3D Debye-H\"{u}ckel free energy (per unit volume)
$\beta \Delta f \sim -\,\kappa_s^3$, the correction to the mean-field
PB free energy (per unit area) scales like
$\beta \Delta f_{pb} \sim -\,\lambda \cdot \lambda^{-3}
\sim  -\,\lambda^{-2}$, in agreement with Eq. (\ref{app3dfree}).
Therefore, the precise calculation leading to Eq. (\ref{3dfree})
justifies the use of the simple picture to illustrate the physics
behind the counterion condensation presented in the Introduction.
We note finally that Eq. (\ref{3dfree}) also contains additional couplings
among the fluctuations of the ``condensed'' and ``free'' counterions.

\section{Results and Discussion}
\label{sec:results}

\begin{figure}
\epsfxsize=4.0in
\centerline{\epsfbox{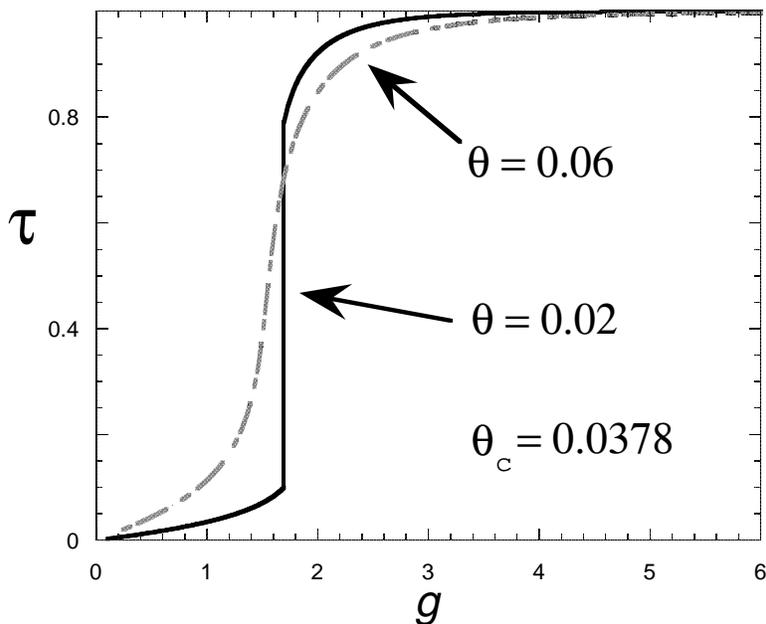}}
\caption{The fraction of
condensed counterions $\tau \equiv Ze n_c/\sigma_0$ as
a function of $g \equiv Z^2 l_B/\lambda$ for different values of
$\theta \equiv {a /(Z^2 l_B)}$. At low surface charge $g \ll 1$,
the counterion distribution is well described by PB theory since
$\tau \ll 1$.  However, at high surface charge, correlation effects
leads to a large fraction of counterion condensed. The condensation
is first-order for $\theta < \theta_c$ and smooth for
$\theta > \theta_c$, where $\theta_c \approx 0.0378$.
The solid line $\theta = 0.02$
corresponds to divalent counterions, where a finite jump occurs at
$g_0 \sim 1.7$ or $\sigma_0 \sim 0.1 \,e\,$ nm$^{-2}$.}

\label{orderparp}
\end{figure}

The central results of this paper follow from the minimization of the total free energy
$f(\tau) = f_{2d}(\tau) + f_{3d}(\tau)$, obtained respectively in Eqs.
(\ref{2D}) and (\ref{3dfree}), with respect to the order parameter $\tau$.
Fig. \ref{orderparp} summarizes the behavior of $\tau$ as a function of
the coupling constant related to the surface charge density $g = Z^2 l_B /\lambda$
and the reduced temperature $\theta \equiv {a /( Z^2 l_B)}$.  For weak coupling $g \ll 1$,
where fluctuation corrections are negligibly small,
the counterions prefer to be free to gain entropy; there
are almost no condensed counterions so that $\tau \approx 0$.
This is not surprising since PB theory is a weak-coupling theory which
becomes exact as $g \rightarrow 0$. However, for higher surface charge density,
where correlation effects become more important,
the behavior of $\tau$ depends crucially on $\theta$.
In particular, for $\theta < \theta_c \approx 0.0378$,
$\tau$ displays a finite jump at $g_0(\theta)$, e.g. $g_0 = 1.695$ at $\theta = 0.02$.
[This corresponds to divalent counterions at room temperature with
$\sigma_0 \sim 0.1 e$ nm$^{-2}$.]  Thus, the system exhibits a first-order
phase transition, in which a large fraction of counterions is condensed (about $80\,\%$).
The physical mechanism leading to this counterion condensation is
the additional binding arising from 2D charge-fluctuations, which dominate
the system at lower temperatures. However, for $\theta > \theta_c$ the behavior
of $\tau$ is completely different; in this regime, there is {\em no phase transition}
and the condensation occurs smoothly.  Thus, the condensation transition
is similar to the liquid-gas transition, which has a line of first-order
transitions terminating at the critical point where a {\em second-order}
transition occurs.  In our case, the critical point is found to be $\tau_c \simeq 0.4$,
$g_c \simeq 1.605$, and $\theta_c \simeq 0.0378$.  Furthermore, if one takes
$l_B \sim 10$ \AA, {\em i.e.} room temperature, and $ a \sim 1\,$\AA,
it follows from the definition of
$\theta$ that there is a critical value of counterion valence $Z_c = \sqrt{a/(l_B \theta_c)}
\simeq 1.62$, below which no first-order condensation transition is possible.
Therefore, divalent counterions behave {\em qualitatively} differently from monovalent
counterions.  In fact, significant differences between mono- and
divalent ions are observed in various biophysical processes.

We stress that fluctuation effects are crucial for this counterion condensation
transition to occur. In fact, it may be viewed as a surface analog of the
bulk transition discussed by Fisher and Levin\cite{fisher}.
These authors predicted a phase separation, where
a strongly correlated, dense phase coexists with a weakly correlated dilute phase
in an ionic system dominated by Coulomb interactions and charge fluctuations.
In our case, the surface breaks the translational symmetry and similar phase separation
occurs in its vicinity.  Indeed, using Eq. (\ref{3dfree})
in the limit of $\tau \rightarrow 0$, {\em i.e.} without assuming the existence of
2D condensate, the system shows a thermodynamic instability at $g \sim 4.4$.
The inclusion of an additional degree of freedom, {\em i.e.}, allowing the counterions
to condense, can only lower the total free energy, suggesting a phase transition
in which the condensate (``liquid'') near the surface coexists
with the more dilute, delocalized counterion (``gas'') distribution.
Indeed, a recent simulation\cite{sim} clearly shows that at
low temperature, most of the counterions reside on the surface,
consistent with our two-fluid picture.  However, our calculation
based on the Gaussian fluctuation theory may break down for very large $g > 10$.
In this regime, a complementary treatment is considered by Shklovskii
in Ref. \cite{shklovskii}, in which the condensed counterions
are assumed to form a 2D Wigner crystal.  That theory also predicts a
strongly reduced surface charge and an exponentially large renormalized
Gouy-Chapmann length, qualitatively similar to our results.  In contrast,
by treating the fluctuations of the condensed and free counterions on an
equal footing, we are able to capture the onset of the condensation (at $g \sim 2$),
which bridges between the regime where PB theory is appropriate, $g \rightarrow 0$,
and the very strong coupling regime, $g \rightarrow \infty$ \cite{shklovskii,sim}.

In summary, we have presented a new mechanism by which the counterions
become condensed so as to neutralize the surface charge of a macroion.
It has been known experimentally that an effective surface charge,
which is always lower than the actual charge, must be introduced
in order to fit experimental data to the PB theory\cite{effective}.
Thus, our theory offers a possible scenario to account for this experimental fact.
In addition, for the case of two highly-charged surfaces,
the PB repulsions between them are greatly reduced due to strong condensation,
and the dominant interaction will be the charge fluctuation
attractions. Thus, this condensation picture may also be crucial to
understanding the like-charged attraction\cite{andy}.
Furthermore, there are some recent experimental\cite{Motschmann}
and simulation\cite{netzaepj2001} indications that are consistent with the
predicted condensation effect.   The experiments\cite{Motschmann}
were performed with a monolayer of cationic surfactant where surface density of the surfactant
and the counterion/salt density are controlled with high accuracy.  The experiments measured a rapid
neutralization (about $90 \%$) of the charged surfactant monolayer by increasing its surface density
(by about $10 \%$).  In some cases, a discontinuous neutralization process is observed\cite{Motschmann}.
Also, recent extensive simulation studies of
uniformly charged surfaces performed in Ref. \cite{netzaepj2001} reports two interesting
observations: First, when $g \gg 1$ there appears a coexistence between two distinct
counterion density distributions:  an exponentially decaying distribution near the immediate
vicinity of the charged surface and an algebraic decaying distribution far away from the the surface.
Note that the exponentially decaying distribution might be associated with our condensed counterions,
which we have assumed to be a delta-function distribution.  Secondly, the specific heat of the simulated system
shows a pronounced hump in the region $10 < g< 100$, though no rigorous proof of the condensation
transition from simulations (and experiments) has been obtained so far.  Indeed, there remains
some fundamental issues to be addressed in the future, for
example, the role of excluded volumes, the discreteness of the surface charge and its mobility,
and higher order (beyond Gaussian) corrections. A recent calculation and simulation shows
that charge discreteness also induces charge localization\cite{dima}. Therefore,
it is possible that these neglected effects may smooth out the first-order transition.
However, we believe that a rapid variation of the condensation
with the surface charge, reflecting the predicted effect, should remain.

\section{acknowledgment}

We would like to thank Ramin Golestanian, Jacob Israelachvili,
Claus Jeppesen, and Kurt Kremer for fruitful discussions.
AL and PP acknowledge support from NSF grants MRL-DMR-9632716, DMR-9624091, DMR-9708646,
and UC-Biotechnology research and Education Program.  DL, PP, and SS
acknowledge support from U.S.-Israel Binational Science Foundation
(BSF) Grant No. 98-00063 and Schmidt-Minerva Center.

\pagebreak


\pagebreak

\begin{thebibliography}{99}
\bibitem{note3}
Note that this approximation neglects (small) contributions from
the coupling among the fluctuations of free and condensed counterions.
\end{thebibliography}
\end{document}